# An Exact and Fast CBCT Reconstruction via Pseudo-Polar Fourier Transform based Discrete Grangeat's Formula

Niloufar Teyfouri, Hossein Rabbani, *Senior Member, IEEE,* Raheleh Kafieh*, Member, IEEE,* Iraj Jabbari

*Abstract*—The recent application of Fourier Based Iterative Reconstruction Method (FIRM) has made it possible to achieve high-quality 2D images from a fan beam Computed Tomography (CT) scan with a limited number of projections in a fast manner. The proposed methodology in this article is designed to provide 3D Radon space in linogram fashion to facilitate the use of FIRM with cone beam projections (CBP) for the reconstruction of 3D images in a sparse view angles Cone Beam CT (CBCT). For this reason, in the first phase, the 3D Radon space is generated using CBP data after discretization and optimization of the famous Grangeat's formula. The method used in this process involves fast Pseudo Polar Fourier transform (PPFT) based on 2D and 3D Discrete Radon Transformation (DRT) algorithms with no wraparound effects. In the second phase, we describe reconstruction of the objects with available Radon values, using direct inverse of 3D PPFT. The method presented in this section eliminates noises caused by interpolation from polar to Cartesian space and exhibits no thorn, V-shaped and wrinkle artifacts. This method reduces the complexity to $O(n^3 \log n)$ for images of size $n \times n \times n$. The Cone to Radon conversion (Cone2Radon) Toolbox in the first phase and MATLAB/ Python toolbox in the second phase were tested on three digital phantoms and experiments demonstrate fast and accurate cone beam image reconstruction due to proposed modifications in all three stages of Grangeat's method.

## I. Introduction

THE growing trend in the use of Computed Tomography (CT) and its application in a wide range of threads, from biology to medicine, and even in the field of industry, adds to the importance of an exact image reconstruction method. Furthermore, increasing the speed of computation is necessary for the next generation real-time imaging systems. By considering Cone Beam CT (CBCT) as one of the most commonly used CTs, we explore an exact and fast volumetric image reconstruction method suitable for conventional geometry of single circular source trajectory.

From 1983 to 1991, Tuy [1], Smith [2] and Grangeat [3] introduced algorithms for an accurate reconstruction of 3D objects in CBCT. All of these approaches include three steps: 1- 3D Radon space generation; 2- Assigning values to the first derivative data in the Radon domain by weighted integral along a line perpendicular to *os*, the distance between the correspondent Radon point on the detectors and origin of coordinate (Fig 1); and 3- Volumetric image reconstruction from Radon data [4]. Inspired by the Grangeat's work, this article will express novelties in all three phases to increase accuracy and speed.

In order to produce 3D Radon space, there are two common types of sampling, including sinogram or linogram. The sinogram sampling [3]–[5] is often used when the reconstruction method, in the third phase, is based on the Filtered Back Projection (FBP) in the time domain, while the linogram fashion [6], [7] is used for Fourier reconstruction algorithms (FRA) in the frequency domain. Of course there are some exceptions like [8] when FFT was applied to the interpolated sinogram in reconstruction stage.

**In the first phase** of this paper, the Radon space will be allocated to a non-Cartesian point set, which is similar to linogram, called pseudo-polar grid [9]. This type of gridding removes artifacts related to interpolation from spherical to Cartesian coordinate system. Then, within a geometric relationship as the rebinning process, a line integration characteristic point, *s* (Fig 1), corresponds to the Radon characteristic point, *ρ*. After rebinning, based on the Grangeat's formula, Radon derivative values are calculated using the linear integration of weighted CBP along the line *t* and then are differentiated with respect to angular vector of *os*.

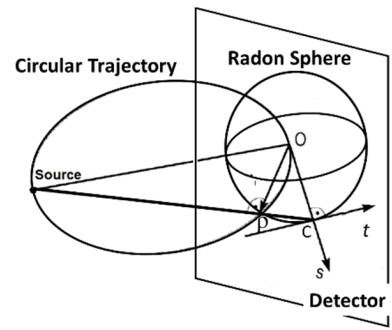

Fig 1. Source-detector arrangement in Grangeat's relation; *o* represents the origin on the detector. The line of Source-*o* is the normal of the detector plane. *ρ* is a point on the Radon sphere that its Radon data is to be calculated. The point *C* is the projection of *ρ* onto the detector. The line *t* is perpendicular to *s* and intersects it at *C*. *t* is the line of integration which will be associated with the Radon transform at point *ρ*.

The integral operation is always reported as a time-

N. Teyfouri, H. Rabbani, and R. Kafieh are with Medical Image & Signal Processing Research Center, School of Advanced Technologies in Medicine, Isfahan University of Medical Sciences, Isfahan 8174673461, Iran. E-mail: rabbani.h@ieee:org.

I. Jabbari is with Department of Nuclear Engineering, Faculty of Advanced Sciences & Technologies, University of Isfahan, Isfahan, Iran.

consuming work in the second step [4].

Table 1
Summary of works done on radial derivative of 3D Radon Generation from CBP

| Author | Time or Frequency Domain | Method | Derivative of Radon Space * |
|---|---|---|---|
| Grangeat et al [3], Lee et al [4], Zhao et al [12]. | Time | Pre- weighting of CBP + Horizontal and vertical derivatives + Rebinning + 1D interpolation + Line integration + Post weighting. | 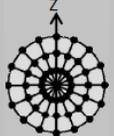 |
| Axelsson et al [13]. | Frequency | Pre- weighting of CBP + Zero padding + 1D FFT + Chirp-z transform + Filtering + Zero filling + Radial inverse FFT + Truncate the zeroes + Put together the linograms + Weighting with $1/cas^2\alpha$ and $1/sin^2\alpha$ + Post-weighting. | 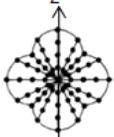 |
| Tam et al [5]. | Time | The Radon data generation directly on the polar grid lines on a set of coaxial vertical planes in the Radon space + only 1-dimensional interpolation in the Radon space. | 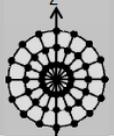 |
| Kudo et al [7] | Frequency | Pre-weighting of CBP + Rotation and bilinear interpolation + FFT + Inverse chirp-z transform + Derivative filtering + Inverse FFT + Post-weighting and rebinning. | 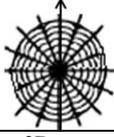 |

*The shapes of this column should be 3D by rotating this area 360 degrees around the vertical axis, Z. For easy demonstration they are shown 2D.

Axelsson et al. [10], proposed an alternative method for integral computation to compensate for the loss of time. They considered the integral as a Radon transformation with respect to Fourier slice theory (FST). Radon values were estimated by radial inverse Fast Fourier transform (FFT) of 2D FFT from an image. Axelsson utilized Chirp-z transform instead of applying FFT in the second stage of 2D FFT, to achieve linogram sampling on the detector. Accordingly, we convert the integral operator to a 2D Radon transformation. However in this paper using rapid 2D DRT [11] based on PPFT, we achieved an increased speed and accuracy in the computation of the integral.

In the derivation stage, Grangeat [3] transformed the derivation in the direction of *s* (Fig 1) into a summation of two components in the horizontal and vertical directions. This action made the appearance of an artifact known as V-shaped [4]. Due to the use of Radon transform instead of integral, we could differentiate in the same direction *s* and thus V-shaped artifact was removed. Differentiation with respect to *s* can also be done in the Frequency domain by multiplying $j\omega$ [13]. Table 1 summarizes the available methods for the computerized generation of the radial derivative of 3D Radon space from CBP.

**In the second phase,** volumetric image reconstruction will be done. We may classify the 3D reconstruction Radon based approaches into two distinct groups according to the Radon domain. Generally these methods are based on FBP and FRA in time and frequency domains respectively.

There are some limitations with exact image reconstruction methods implemented in the time domain, the most significant of which pertains to noise and artifacts. Three types of artifacts associated with the Grangeat's formula are reported; they are referred to as thorn, wrinkle, and V-shaped artifacts [4].

Thorn is caused by backprojection of the second derivative of the Radon data on the meridian planes. The proposed algorithm in this paper applies 3D inverse DRT to the Radon data in the reconstruction stage without any additional derivative of the Radon data. It will be shown that the image is free of this type of noise using our proposed method. The second derivative of the Radon data is precisely the reason of wrinkle and V-shaped artifact appearance that would be also eliminated by our proposed algorithm. Another group of artifacts stem from the discrete nature of numerical implementation. Lee et al [4] displayed these artifacts in their implementation of the Grangeat's formula. Providing a successful method to remove this type of artifacts and therefore a fast reconstruction of an object from continuous projections is the objective sought by this article.

In the present work, Direct Fourier reconstruction Method (DFM), which potentially enjoys high speed implementation in the frequency domain, has been selected. DFM is closely linked to the FST which states that the 1D FFT of a projection at angle $\theta$ is a radial slice through the 2D FFT of the object at direction $\theta$. Accordingly, DFM is composed of following steps: applying FFT to padded projections; interpolation of 2D Cartesian FFT grid from the polar grid; and reconstruction of the image by a 2D inverse FFT [14].

In FRAs, a major challenge is the provision of data on spherical coordinates by the FST. That is while FFT inversion requires the data in Cartesian coordinates. Meanwhile, the frequency domain needs accurate interpolation in order to convert the spherical to Cartesian lattice.

Table 2
Summary of Works Done on Volumetric Image Reconstruction from Radial Derivative of 3D Radon.

| Author | Time or Frequency Domain | Method |
|---|---|---|
| Grangeat et al [3], Lee et, al [4]. | Time | FBP: Derivative filtering of derivative Radon data in the radial direction + 2nd derivatives of Radon data on a meridian plane + backprojection on a meridian plane + 2nd backprojection on an axial plane. |
| Tam et, al [15]. | Time | FBP of the Radon transform of the object residing on a set of coaxial planes + FBP operates on parallel beam projection images of the object on the same set of coaxial planes. |
| Dusaussoy et al [14]. | Frequency | Calculation of the concentric cubes + Interpolation along the sides of concentric rectangles to calculate a Cartesian grid in each meridian plane + Interpolation along the sides of concentric squares to calculate a Cartesian grid in each horizontal plane + 3D inverse FFT |
| Axelsson et al [6]. | Frequency | Sampling the Radon space on a number of vertical planes in a modified 2D linogram fashion (as a result of the interpolation) + Reconstruction of the vertical planes with the 2D linogram method + Filling up Fourier cube + Reconstruction of the horizontal planes directly with the linogram. |
| Schaller et, al [8]. | Frequency | FFT along radial lines in the Radon space + Interpolation from the spherical to a Cartesian grid using a 3D gridding step in the frequency domain + 3D inverse FFT. |
| Zhao et al [12]. | Frequency | Filtered backprojection in Fourier domain: Calculates the contributions to the three dimensional Fourier transform of the 3D object by derivative of Radon related to each projection separately + Add up all contributions from all projections in Fourier domain + Average the contributions from source geometries + 3D Inverse FFT. |
| Kudo et al [7]. | Frequency | Apply radial FFT on the derivative Radon sampling + First derivative filtering + 2D Chirp z-transform + Inverse FFT + Rotation and trilinear interpolation + addition of component images |

Numerical performance of this task is difficult when there is only a finite number of samples available. The reason is that the slightest error in the frequency domain would have impacts on the entire image [14].

A review of surveys conducted on implementing this phase is available in Table 2. CT reconstruction from CBP is a natural extension of the 2D case. Already, Averbuch et al. [11] reconstructed 2D images by applying 2D inverse DRT on the 2D Radon data obtained from fan beam projections. Our research demonstrates the ability of 2D and 3D DRT in the generation of 3D Radon data from CBP in a fast and exact manner. Moreover, inverse 3D DRT is a powerful tool for volumetric image reconstruction from 3D Radon data. Compared with the reviewed works, the main advantages of this paper are as follows:

- o Discretization and verification of the Grangeat's formula which works well for continuous objects and detectors with non-unit pixel size.
- o Fast implementation of the Grangeat's formula.
- o No noise due to discretization and reconstruction method.
- o Proof of applicability of inverse 3D DRT to reconstruct a volumetric object from continuous projections.

This paper is structured as follows. In Section II, the theoretical background of 2D and 3D FST and DRT are described. This section provides all necessary details to understand the implementation of the Grangeat's formula. Section III involves different stages of the method including step-by-step implementation and verification of the Grangeat's formula. In Section IV, the criteria and data used to evaluate the algorithm will be expressed. Section V is then dedicated to experiments. We show the results obtained from computer simulations to illustrate the performances of the formula. Discussions and conclusions are given in Sections VI and VII respectively.

## II. PRELIMINARIES

In this section, we review some theories and techniques that will be used in the next sections to develop the exact reconstruction method. We provide an overview of FST, PPFT and rapid DRT in 2D and 3D.

### A. 2D FST

According to 2D FST, the 1D FFT ($F_1$) of the parallel projection ($P_l$) of a 2D image $I(x, y)$ in a direction equals the slice ($S_l$) of the 2D FFT ($F_2$) of $I$ in the same direction (Fig 2).

$$F_1 P_l = S_l F_2 \quad (1)$$

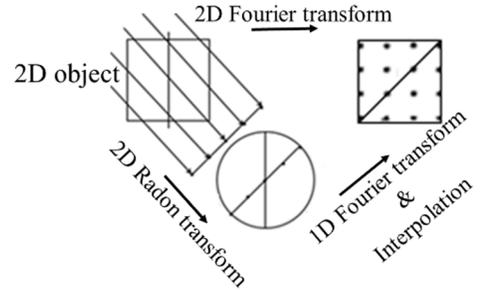

Fig 2. Illustration of 2D FST

### B. 3D FST

According to 3D FST, the radial Fourier transform ($F_1$) of a projection ($P_l$) of a 3D image $I(x, y, z)$ in a direction, i.e. the Fourier transform of data along a line through the origin in the Radon space, equals an 1D slice ($S_l$) of the 3D Fourier transform ($F_3$) of $I$ in the same direction (Fig 3).

$$F_1 P_l = S_l F_3 \quad (2)$$

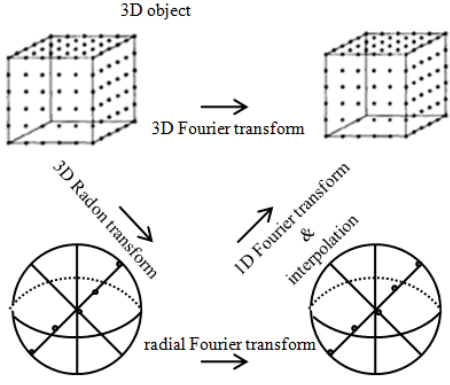

Fig 3. Illustration of 3D FST.

## C. Pseudo Polar Fourier Transform (PPFT)

The spherically sampled Fourier space achieved by 3D sinogram sampled Radon space requires 3D-interpolation to obtain the 3D FFT of the object.

Pseudo-polar or linogram sampling would be beneficial in removing the interpolation stage. Two-dimensional linogram sampling of 2D Radon space delivers samples on concentric quadrates in 2D Fourier space (Fig 4).

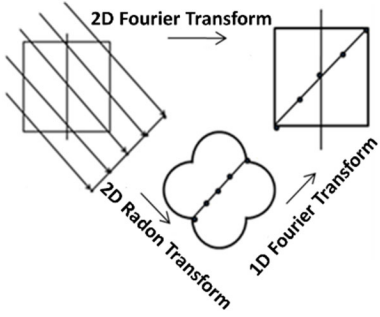

Fig 4. 2D PPFT

Generalization of the Fourier quadrate to 3D is the Fourier cube (Fig 5). PPFT evaluates the Fourier transform on a non-Cartesian pseudo-polar grid. The rapid exact evaluation of the Fourier transform at these non-Cartesian grid points is possible using the fractional Fourier transform or chirp z-transform [16].

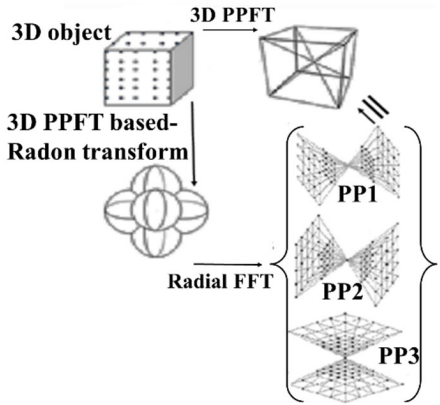

Fig 5. 3D PPFT.

The set $(P \triangleq PP_1 \cup PP_2 \cup PP_3)$ (Fig 5) is called the pseudo-polar grid [16].

### D. Rapid 2D DRT, $\Re_2$.

We consider a 2D continuous object, $I(u, v)$ as $n \times n$ image with the pixel size of $du$ in both directions of $u$ and $v$ axis, such that $I(u, v)=0$ outside $[-su/2, su/2] \times [-su/2, su/2]$ and $su = n \times du$.

A 2D DRT is defined by summing the interpolated samples of $I$ lying on lines with absolute slope less than one (3).

There are two types of lines. The first is the basically horizontal line having the form of $v = qu + p$, where $|q| \leq 1$, (Fig 6.a) and the second is the basically vertical line that is a line of the form $u = qv + p$, where $|q| \leq 1$ (Fig 6.b).

$$u = (-n/2, \ldots, n/2 - 1) \times du, \tag{3}$$
$$v = (-n/2, \ldots, n/2 - 1) \times du, \tag{4}$$
$$p = (-n, \ldots, n) \times du, \tag{5}$$
$$q = \left(-\frac{n}{2}, \ldots, \frac{n}{2}\right)/(n/2). \tag{6}$$

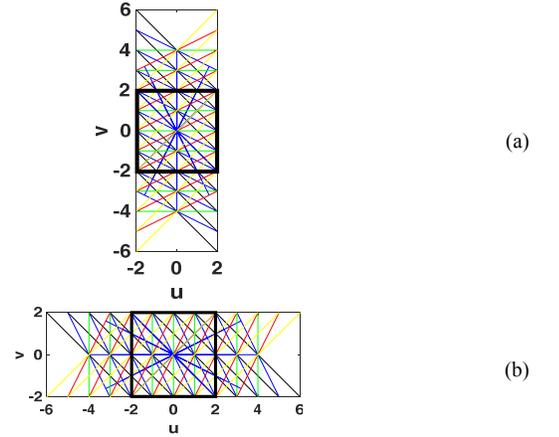

Fig 6. The black square is 2D image $I(u, v)$ with the pixel number of 4 and $du = 0.5$. (a) Basically horizontal lines for n = 4. (b) Basically vertical lines for n = 4.

Finally, the notion used for 2D DRT is $\Re_2$.

$$\Re_2(q, p) = \begin{cases} R_1 = Radon(\{v = qu + p\}, I): horizontal\ lines \\ R_2 = Radon(\{u = qv + p\}, I): verticaal\ lines \end{cases} \tag{7}$$

2D FST proves the relation between the 2D DRT, and the 2D discrete PPFT, $PP_i$, corresponding to image I, as shown in (8). $F_1^{-1}$ is the 1D inverse FFT and $o$ is the operator.

$$R_i I = F_1^{-1} o\, PP_i I \quad (i = 1,2) \tag{8}$$

The set $\Re_2(q, p) \triangleq R_1 \cup R_2$ is called 2D DRT.

### E. Rapid 3D DRT, $\Re_3$.

Inspired by the definition of the 2D DRT given in section II.D above, the 3D discrete Radon transform is defined by summing the interpolated samples of a continuous 3D image $I(u, v, w)$ (with pixel size of du in directions of $u$, $v$ and $w$ axis) lying on planes with certain constraints. The available planes may be considered in three types. The equation of the planes is in accordance with (9).

$$x = q_1 y + q_2 z + p, \quad (9)$$
$$y = q_1 x + q_2 z + p, \quad (10)$$
$$z = q_1 x + q_2 y + p. \quad (11)$$

Considering the first set of planes, the following relations are proposed for 3D DRT in the x- direction, ($\Re_{31}$).
$$Radon(\{x = q_1 y + q_2 z + p\}, I) = \Re_{31}(q_1, q_2, p). \quad (12)$$
Where
$$x = (-n/2, \ldots, n/2 - 1) \times dx, \quad y = x, \quad z = x, \quad (13)$$
$$p = (-3n/2, \ldots, 3/2n) \times dx, \quad (14)$$
$$q_1 = \left(-\frac{n}{2}, \ldots, \frac{n}{2}\right) / \left(\frac{n}{2}\right), \quad q_2 = q_1 \quad (15)$$
$$u = \left(-\frac{n}{2}, \ldots, \frac{n}{2} - 1\right), \quad v = u, \quad w = u \quad (16)$$

Similarly, the 3D DRT referred y-, z-direction are defined as (17) and (18):
$$Radon(\{y = q_1 x + q_2 z + p\}, I) = \Re_{32}(q_1, q_2, p). \quad (17)$$
$$Radon(\{z = q_1 x + q_2 y + p\}, I) = \Re_{33}(q_1, q_2, p). \quad (18)$$
The notion used for 3D DRT is $\Re_3(q_1, q_2, p)$.
$$\Re_3(q_1, q_2, t) \triangleq (\Re_{31} \cup \Re_{32} \cup \Re_{33}) \quad (19)$$

## III. METHOD

Grangeat [3] proposed a formula that links CBP on virtual detector to 3D Radon data. In this paper, the discrete version of this method is optimized by several fast and accurate algorithms. We demonstrate the applicability of the inverse DRT for the reconstruction of a 3D object from continuous projections.

### A. The Grangeat's Formula

The Grangeat's Formula can be expressed as follows:
$$\frac{\partial}{\partial \rho} Rc(\rho, \theta, \varphi) = \frac{1}{\cos^2 \beta} \frac{\partial}{\partial s} \int_{-\infty}^{+\infty} \frac{SO}{SA} X_\psi(s, \alpha) dt \quad (20)$$

$Rc$ is the continuous Radon transform defined in (21).
$$Rc(\rho, \theta, \varphi) = \iiint_{-\infty}^{+\infty} I(x, y, z) \, \delta(x \sin \theta \cos \varphi \quad (21)$$
$$+ y \sin \theta \sin + z \cos \theta - \rho) dx dy dz$$

$I(x, y, z)$ is the 3D object in the Cartesian coordinate. *SO* signifies the distance between the source *S* and the origin *O*, *SA* denotes the distance between the source and an arbitrary pixel of $X_\psi$ along *t*. Based on (20), each characteristic point of the 3D Radon space C:$(\rho, \theta, \varphi)$ (Fig 7 a) is mapped to a pixel of a detector at angle $\psi$ from the projection D:$(s, \alpha, \psi)$ (Fig 7.b), where OC line is perpendicular to integration plane (Fig 7.a).
$X_\psi$ is the virtual detector at angle $\psi$ of projection. Fig 8 shows that each detector located at a constant distance of SP from source, *S*, is convertible to virtual detector by multiplying its side length at (*SO/SP*). It's clear that there is no difference between the pixel number of the detector and its virtual correspondent.

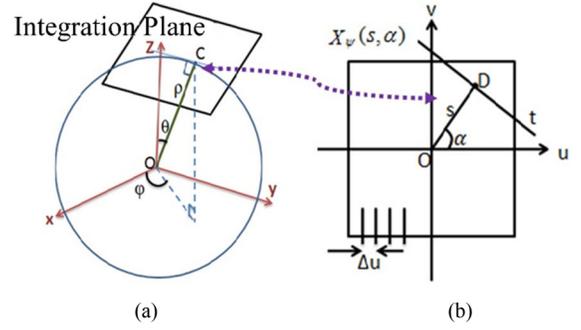

Fig 7. (a) 3D Radon space on sphere coordinate. (b) Detector on polar coordinate.

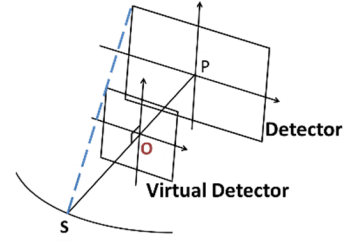

Fig 8. Virtual detector in CBCT

As Fig 9 shows, $X_\psi(s, \alpha)$ is the detector value a distance of *s* away from the center of detector *O* along the line *t* in a perpendicular position to *OD*. *OC* is perpendicular to *SD*. $\beta$ is the angle between the line *SO* and SD.

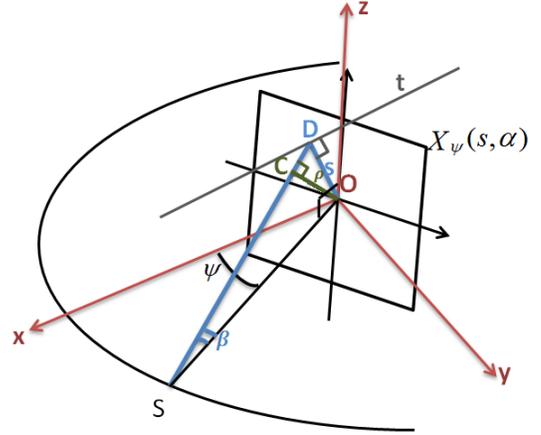

Fig 9. Relationship between a characteristic point of 3D Radon space, C, and its mapping, D, on the virtual detector

### B. Mapping 3D Radon Space on detectors

Assuming that all the integral planes passing through the object *f(x)* cross the X-ray source, S, located at the angle of $\psi$ with x- axis, the locus of the normal vector to these planes will be located on a sphere known as the Radon shell of diameter *SO* [10] (Fig 10).

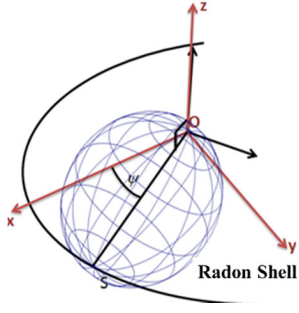

Fig 10. Relation between Radon shell and spherical coordinate

The equation of Radon shell at angle $\psi$ is in accordance with (22).

$$(x - x_c)^2 + (y - y_c)^2 + (z - z_c)^2 - r^2 = 0. \qquad (22)$$

$$r = \frac{SO}{2}, \qquad (23)$$

$$(x_c, y_c, z_c) = \left(\frac{SO}{2}\cos\psi, \frac{SO}{2}\sin\psi, 0\right). \qquad (24)$$

$$x^2 + y^2 + z^2 = SO\,(x\cos\psi + y\sin\psi). \qquad (25)$$

Considering the characteristic point of the 3D Radon space C in sphere coordinates,

$$x^2 + y^2 + z^2 = \rho^2 = SO\,(x\cos\psi + y\sin\psi). \qquad (26)$$
$$x = \rho\sin\theta\cos\psi, \qquad (27)$$
$$y = \rho\sin\theta\sin\psi, \qquad (28)$$
$$z = \rho\cos\theta. \qquad (29)$$

The geometric relation between C: $(\rho, \theta, \varphi)$ and D: $(s, \alpha, \psi)$ will be defined in (30)- (32)

$$s = \frac{\rho\,SO}{\sqrt{SO^2 - \rho^2}} \qquad (30)$$

$$\alpha = \tan^{-1}\left(\frac{\theta}{|\theta|} \frac{1}{\sqrt{\frac{SO^2 - \rho^2}{SO^2(\cos\theta)^2} - 1}}\right) \qquad (31)$$

$$\psi = \varphi - \left|\cos^{-1}\left(\frac{\rho}{SO\,\sin\theta}\right)\right| \qquad (32)$$

Fig 7 and Fig 9 clearly demonstrate this relation. The triangle of Fig 11 should be established in each source position.

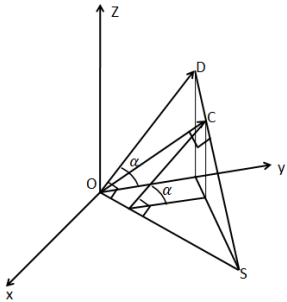

Fig 11. The geometric relation between a Radon characteristic point, C and correspondent detector pixel, D.

Therefore, using an interpolation step, all Radon space points are mapped to the points on the detectors.

Due to the timing of the interpolation command in the MATLAB programming language, this part of the program was written in Python, and then the output was transmitted to MATLAB.

*C. The Discrete Grangeat's formula*

The goal of this paper is to utilize 2D and 3D DRT to improve the speed and accuracy of the discrete version of the Grangeat's formula [3]. Equation (33) is the discrete form of (20).

$$\Delta\Re_3(I(\rho,\theta,\varphi)) = \frac{SO}{(\Delta\rho)\times(\Delta m)^3\times\cos^2\beta}\,\Delta\Re_2\left(\frac{X_\psi(s,\alpha)}{SA}\right)\Delta t \qquad (33)$$

where the pixel size of the cubic object is $\Delta m$, and $\Delta\rho$ is the distance between two consecutive $\rho$s in the diameter direction $(\theta, \varphi)$ of 3D Radon linogram sampling. $SO$ is the distance between the source $S$ and the origin $O$. $\Delta\Re_2\left(X_\psi(s,\alpha)\right)$ and $\Delta\Re_3(I(\rho,\theta,\varphi))$ are the difference between two consecutive (with a same angular direction and consecutive radius) 2D and 3D DRT described in II.D and II.E sections respectively. Averbuch et al. elaborate $\Re_2$ [17] and $\Re_3$ [16] in Cartesian coordinates, but in this paper they are used in polar and spherical coordinates respectively (as described in D and E of this section).

The line integration in (20) in the direction of t-axis (Fig 7.b) is just the sum of pixels of detector in the same direction. So it is convertible to 2D Radon transform. The line equation related to $t$ is defined as horizontal and vertical lines in (7). In that equation $q$ is determined by the location of $(s, \alpha)$ mapped from each $(\rho, \theta, \varphi)$. In detectors with the pixel size of $\Delta u$, the distance between pixels along the line, $\Delta t$, is computed by $\sqrt{\Delta u^2 + \Delta v^2}$. For example in the horizontal line, $\Delta v = q \times \Delta u$. So,

$$\Delta t = \sqrt{(\Delta u)^2 + (q\times\Delta u)^2} \qquad (34)$$

$SA$ denotes the distance between the source and an arbitrary pixel of $X_\psi$, $A$, along $t$. For example if $(s,\alpha)$ is located on horizontal lines, then

$$SA = \sqrt{u^2 + v^2 + SO^2} \qquad (35)$$
$$= \sqrt{u^2 + (qu+p)^2 + SO^2}$$

If $SO\to\infty$, it can be said $SA = SO$, otherwise it is a very time-consuming procedure to calculate $\frac{X_\psi(s,\alpha)}{SA}$ for all $q$ and $p$.

*D. 2D Radon Sampling*

The notion used by Averbuch et al. [17] for 2D Radon transform is based on slopes, $q$ and intercepts, $t$ in (7) to specify a line. However it is conventional to use the distance from the origin, $s$ and the angular direction, $\alpha$, to clear the normal vector of line. Equations (36)-(39) describe converting the Cartesian coordinates to polar based on basically horizontal and vertical lines integration. For basically horizontal lines we would have:

$$s = |p|/\sqrt{1+q^2} \qquad (36)$$
$$\alpha = \tan^{-1}\left(\frac{-1}{q}\right) \qquad (37)$$

and for basically vertical lines:

$$s = |p|/\sqrt{1+q^2} \qquad (38)$$
$$\alpha = \tan^{-1}(-q) \qquad (39)$$

Fig 12 shows the position of pseudo-polar based grid or linogram 2D sampling of the Radon space on the detectors.

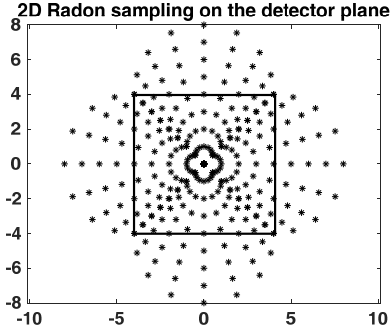

Fig 12. 2D Radon sampling determining the position of normal vector to integration line on the detector, black square

*E. 3D Radon Sampling*

The definition of 3D DRT described by Averbuch et al [16] uses slopes, $(q_1, q_2)$, and intercepts, $(p)$, based on (9) and (13) to designate a specific plane which crosses the object. This notation for the Radon space is less common. Usually, a normal vector, $(\rho, \theta, \varphi)$, in spherical coordinate used to elaborate the position of Radon characteristic point, (Fig 7.a). $\rho$ is the distance of the plane from the origin and the angle of $\theta$ and $\varphi$ are indicated in Fig 7.a.

The range of $(\rho, \theta, \varphi)$ which defines all x-planes, satisfies (40)-(42).

$$\rho = |p|/\sqrt{1 + q_1^2 + q_2^2} \tag{40}$$

$$\theta = \tan^{-1}\left(\sqrt{1 + q_1^2}/q_2\right) \tag{41}$$

$$\varphi = \tan^{-1}(-q_1) \tag{42}$$

Similarly, for y-planes, we have:

$$\rho = |p|/\sqrt{1 + q_1^2 + q_2^2} \tag{43}$$

$$\theta = \tan^{-1}\left(\left(\sqrt{1 + q_1^2}/q_2\right)\right) \tag{44}$$

$$\varphi = \tan^{-1}(-1/q_1) \tag{45}$$

and for z-planes:

$$\rho = \frac{|p|}{\sqrt{1 + q_1^2 + q_2^2}} \tag{46}$$

$$\theta = \tan^{-1}\left(-\sqrt{q_1^2 + q_2^2}\right) \tag{47}$$

$$\varphi = \tan^{-1}(q_2/q_1) \tag{48}$$

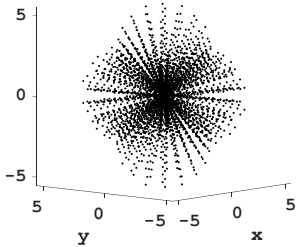

Fig 13. 3D Radon sampling in spherical coordinates determining the position of the normal vector to integration planes on the 3D object in time domain

Fig 13 and Fig 14 show the position of pseudo-polar based 3D sampling of the Radon space in time and the frequency domain respectively.

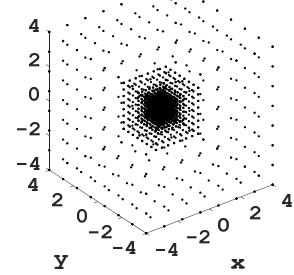

Fig 14. 3D Radon sampling in the spherical coordinate determining the position of normal vector to the integration planes on the 3D object in frequency domain

*F. Numerical Integration of the Derivative of the Radon Transform*

After computing $\Delta\mathfrak{R}_3$, numerical integration was performed to calculate $\mathfrak{R}_3(\rho, \theta, \varphi)$ via the trapezoidal method. This method approximates the integration over an interval by breaking the area down into trapezoids with more easily computable areas. For integration with N+1 evenly spaced points, the approximation is defined as:

$$\int_a^b f(x)dx = \frac{1}{2}\sum_{n=1}^{N}(x_{n+1} - x_n)[f(x_n) + f(x_{n+1})] \tag{49}$$

$(x_{n+1} - x_n)$ is the spacing between each consecutive pair of points.

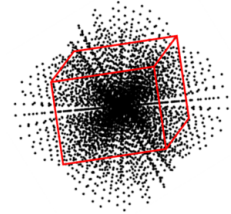

Fig 15. Both ends of all Radon diameters are outside the object.

As Fig 15 shows, the first point with any diameter of the Radon space is outside the support domain of the object. Hence, the initial value of the integral in each diameter is zero. This event is due to use of linogram sampling or pseudo-polar grid.

*G. Compensation weighting factor $(1/\Delta\rho)$ to radial Radon Transform in the linogram fashion*

$\Delta\rho$ represents the sample distance for radial line in the 3D Radon space. Denser sampling indicates that more plane-integrals penetrate the object and that the sum of radial line becomes higher than that of a more sparsely sampled line [10]. Now we have to make up for this difference by multiplying each radial line by $(1/\Delta\rho)$ in equation (33); otherwise, the denser sampled radial lines will show larger influence.

## H. Shadow zone

Cone-beam consistency conditions are mathematical relationships between different CBP, and therefore they describe the redundancy or overlap of information between projections [18].

Necessary consistency conditions are any description of the redundancy in ideal projection data. They result from mathematical relationships between the unknown object and its projections. Sufficient consistency conditions are required to ensure that they are compatible with some object functions. We say that a set of conditions is *full* if the conditions are both necessary and sufficient.

One of the most important sufficient conditions states that one can reconstruct the object if on every plane intersecting the object there exists at least one cone-beam source point.

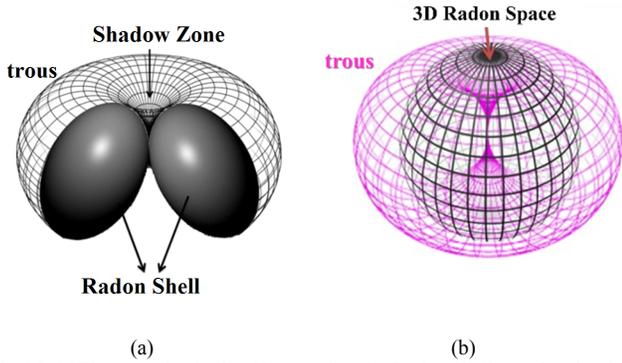

(a)             (b)

Fig 16. (a)The torus is obtained by moving the Radon shells on the circular source path. (b) Display of shadow zone in the center of torus.

In this study, a single circular source trajectory was adopted. This trajectory does not satisfy the sufficient condition for exact reconstruction. As a result, there exists a *shadow zone* in the Radon domain, as shown in Fig 16. This zone is defined by

$$|\rho| > SO|\sin\theta| \qquad (50)$$

With only a single circular orbit, we must fill in this shadow zone, where linear interpolation was selected for this filling.

## I. Volumetric Image Reconstruction

Averbuch et al [11] showed that 3D DRT is invertible. Given the 3D DRT $\Re_3(\rho, \theta, \varphi)$ on PP grid, it is possible to uniquely recover volumetric image, I(x, y, z) via fast direct inversion of the 3D PPFT (Fig 17) [19]. The pseudo code of the proposed recovery is shown in Algorithm 1.

Algorithm 1: Algorithm of rapid 3D inverse DRT for volumetric image reconstruction from 3D linogram sampling.

---
Initialize: axis=[x, y, z] for i=1:3{
- For each k, l and j, $\hat{I}_{[i]}(k,l,j) = $ 1D DFT on $Radon_{[i]}(k,l,j)$ [20]}
- Direct Inversion of the 3D PPFT on $\hat{I}$ [19]:{
  - ✓ Resampling the pseudo-polar grid to an intermediate Cartesian grid.
  - ✓ Recovering I from the samples of $\hat{I}$.}
---

To reconstruction an image of size $n \times n \times n$, Radon space, $Radon_{[i]}(k, l, j)$, will be a four dimensional array of size $3 \times (3 \times n+1) \times (n+1) \times (n+1)$. For example k, l, j indices in $Radon_{[1]}(k, l, j)$ mean pseudo-radius (unit steps) in x-direction, pseudo-angle in y-direction and pseudo-angle in z-direction, respectively.

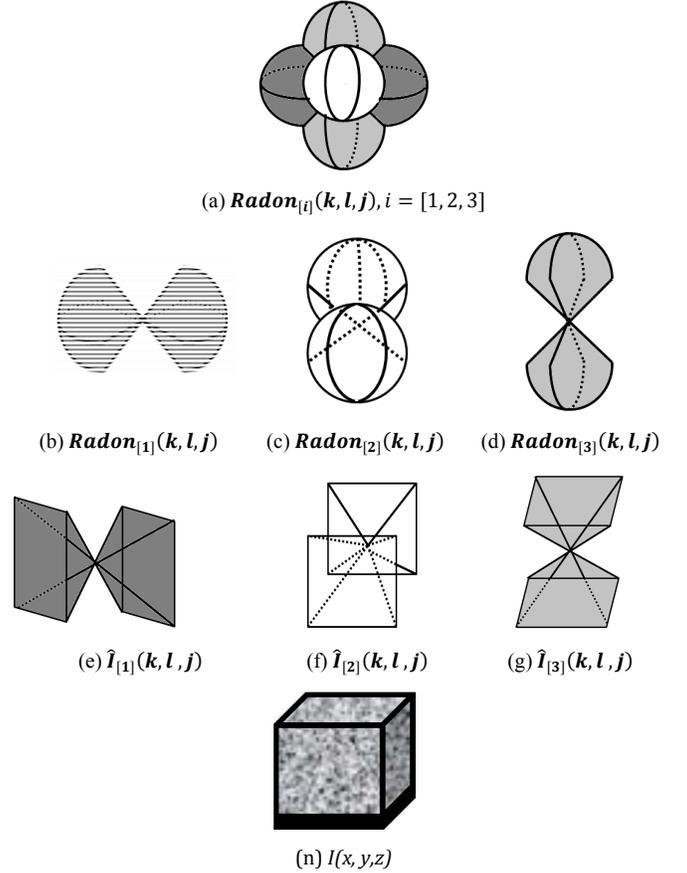

(a) $Radon_{[i]}(k,l,j), i = [1,2,3]$

(b) $Radon_{[1]}(k,l,j)$    (c) $Radon_{[2]}(k,l,j)$    (d) $Radon_{[3]}(k,l,j)$

(e) $\hat{I}_{[1]}(k,l,j)$    (f) $\hat{I}_{[2]}(k,l,j)$    (g) $\hat{I}_{[3]}(k,l,j)$

(n) $I(x,y,z)$

Fig 17. The process of recovering 3D object from 3D Radon space.

## J. Computational Complexity

One of the overriding motives for this work is to find faster (Fourier) methods for reconstruction. In order to evaluate the computational complexity of our method, the total number of Floating Point Operations (FLOPs) is estimated. Addition, subtraction, multiplication and division are considered each as one FLOP in our estimations.

Assume a $N \times N \times N$ volume is to be reconstructed, the detector size is $N \times N$ and we use $M$ projections.

FDK is a generalization of the popular filtered backprojection method for the 2D case and its complexity is $O(MN^3)$ or $O(N^4)$ since $M = O(N)$ [6], [21].

As mentioned in section I, the process of image reconstruction from CBP involves two phases of Radon space and then image generation. The computational complexities for the various reconstruction steps are as follows:

1. 1st phase: Radon Generation.
- Pre-weighting: The cost for the pre-weighting of $N \times N$ detector values in M projections is $MN^2$ FLOP, then the time complexity of pre-weighting is $O(N^3)$.
- 2D DRT: The cost for computing 2D DRT ($\Re_2$) for each $N \times N$ detector is $O(N^2 \log N)$ operations [17],

and for all detectors is $O(MN^2 \log N)$. So, the total complexity of $\Re_2$ is $O(N^3 \log N)$.

- Differentiation in the s direction: The cost for one line numerical gradient which involves $N$ points and requires $2(N-1)$ subtractions in the numerator and denominator, $(N-1)$ divisions and $(N-1)$ additions is $4(N-1)$ FLOP. Each detector includes $2(N+1)$ lines with $(2N+1)$ points. So, $2(N+1) \times 8NM$ FLOP yields the $O(N^3 \log N)$ as the time complexity of the derivative in s directions.
- Post-weighting: The cost for the post-weighting of $2(N+1)(2N+1)$ points in $M$ detector plane is $2M(N+1)(2N+1)$ FLOPs or $O(N^3)$ operations.
- Interpolation: The desired Radon data points for the reconstruction are positioned on PP grid, which do not coincide exactly with the data on a Radon shell of any source position. Therefore, interpolation between the two closest Radon shells is necessary. In our experiments, we have used tri-linear interpolation, which consists of seven linear interpolations (Fig 18).

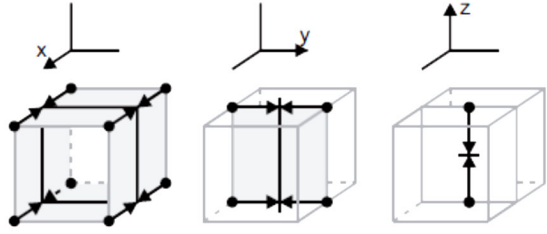

Fig 18: Tri linear interpolation.

Equation (51) is understood by Fig 19 related to linear interpolation.
$$f(x_{n+dx}) = (f(x_{n+1}) - f(x_n)) \times dx + f(x_n) \qquad (51)$$

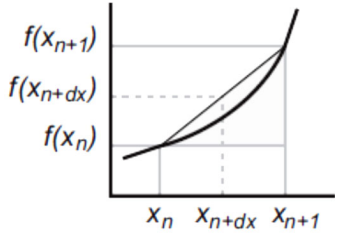

Fig 19: Linear interpolation plot.

Thus for one linear interpolation one subtraction, one multiplication and one addition are required, which makes 3 FLOP for one linear interpolation and 7*3 = 21 FLOP for one tri-linear interpolation. Considering the $3 \times (3 \times n+1) \times (n+1) \times (n+1)$ Radon data points, the cost for this steps is $21 \times (3 \times n+1) \times (n+1) \times (n+1)$ FLOPs or $O(N^3 \log N)$ operations.

- Computation of Radon values from derivatives: Based on trapezoidal numerical integration, the integral of $3 \times (n+1) \times (n+1)$ lines with $(3 \times n+1)$ points includes $3 \times (3 \times n) \times (n+1) \times (n+1)$ summations, $3 \times (n+1) \times (n+1)$ multiplications and $3 \times (n+1) \times (n+1)$ subtractions. The total time complexity is $O(N^3 \log N)$.

2. 2nd phase: Image reconstruction from 3D Radon space.

- Resampling: a total of $O(N^3 \log N)$ operations are needed for the resampling step [19].
- Recovering the 3D image from samples of $\hat{I}$ takes $O(N^3 \log N)$ operations [19].

In short, we proposed an $O(N^3 \log N)$ method for 3D reconstruction from CBPs.

### K. Flowchart of the proposed method

We demonstrate the complete process using a block diagram in Table 3.

Table 3
Block Diagram of the Complete Process

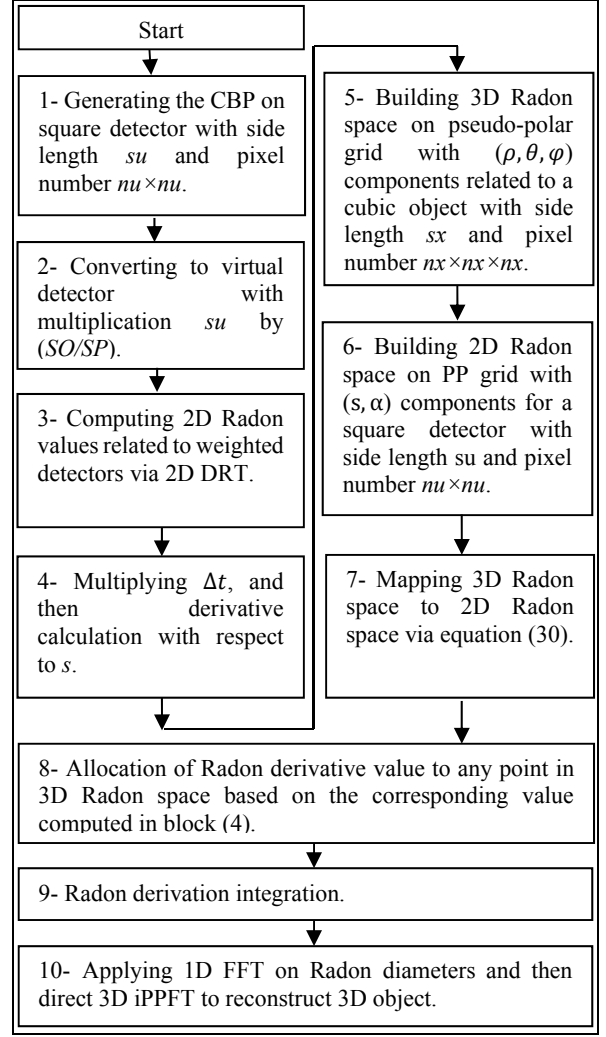

### IV. MATERIALS AND EVALUATION INDEXES

Three synthetic phantoms are utilized to validate the proposed method: The Shepp-Logan [22], Compressive Sensing (CS) phantom [23] and Zubal Head [24].

Fig 22 shows a representative slice of three mentioned phantoms. The digital phantom was projected by CBCT geometry with single circular source trajectory using TIGRE toolbox [24].

In all stages of computerized implementation, we assume that the object and the detector are continuous and their pixel sizes are not unit. So, the customizable parameters are the

following:
- The length of each side of the cubic object in mm: *sx*.
- The voxel number of the object: *nx×nx×nx*.
- The length of each side of the real detector located at the distance of SP in Fig 8 in mm: *su×su*.
- The pixel number of the detector: *nu×nu*.
- The distance between the source and the real detector in mm: SP.
- The distance between the source and the middle plane of the object in mm: SO

Details of the different phantoms are summarized in Table 4. Considering the last paragraph of subsection III.C, to save the run time, SO is considered a large value.

Table 4
Customized parameters in two selected phantoms

| No | Phantom | sx | nx | su | nu | SP | SO |
|----|---------|----|----|----|----|----|----|
| 1 | Shepp-Logan | 16 | 32 | 128 | 256 | 1500 | 1000 |
| 2 | Shepp-Logan | 64 | 64 | 128 | 256 | 1500 | 1000 |
| 2 | Head | 48 | 64 | 128 | 256 | 1500 | 1000 |
| 3 | CS | 32 | 64 | 128 | 256 | 1500 | 1000 |

To quantitatively evaluate the quality of reconstructed images, the peak signal-to-noise ratio (PSNR), contrast-to-noise ratio (CNR), structural similarity (SSIM) [25] are selected as performance metrics.

PSNR is defined as:

$$PSNR = 10 log_{10} \frac{\mu_{max}^2}{MSE} \quad (52)$$

Where MSE is the mean-squared error between the reconstructed image and the reference image and $\mu_{max}$ is the maximum possible value of the image.

CNR is defined as:

$$CNR = |\mu_{ROI} - \mu_{Ref}| / \sqrt{\sigma_{ROI}^2 + \sigma_{Ref}^2} \quad (53)$$

Where $\mu_{ROI}$ and $\mu_{Ref}$ are the mean intensity of a selected ROI and the reference uniform background region, $\sigma_{ROI}$ and $\sigma_{Ref}$ are the standard deviation of the ROI and the reference region.

The SSIM metric is defined as:

$$SSIM(a,b) = \frac{(2\mu_a \mu_b + C_1)(2\sigma_{ab} + C_2)}{(\mu_a^2 + \mu_b^2 + C_1)(\sigma_a^2 + \sigma_b^2 + C_2)} \quad (54)$$

Where *a* and *b* are two local windows of size 8 × 8 pixels in two images with the same position. $\mu_a$ and $\sigma_a$, $\mu_b$ and $\sigma_b$ are mean and standard deviation in each window, respectively. $\sigma_{ab}$ is the covariance between the two windows. $C_1$ and $C_2$ are two constants to avoid instability. In this study, $C_1$ and $C_2$ were chosen as:

$$C_1 = (0.01 \mu max)^2 \text{ and } C_2 = (0.03 \mu max)^2 \quad (55)$$

SSIM is used to measure similarity in the structure between the two windows. As the two windows move pixel-by-pixel over the reconstructed image and the reference image, we obtain a SSIM map. In practice, we use a single Mean-SSIM (MSSIM) value to evaluate the overall image quality by simply averaging SSIM values.

## V. RESULTS

### A. Quantitative Comparison

In this section, the performance of the proposed method is assessed by expressing the values of evaluation indexes. In this appraisal, the impact of some parameters on the reconstructed image quality is determined by the number of detector pixels, and the type of interpolation in shadow zone. In order to express the impact of the samples of continuous projections after conversion to discrete projections, Table 5 shows a fixed length detector with the different choice of pixel number (360 is the number of projections). So, 3D DRT and its inverse are able to model accurately the continuum in parallel with the increase in the number of samples.

Table 5
Comparing different pixel numbers in a detector with *su=64* related to CS phantom

| Pixel Number | PSNR | CNR | SSIM |
|---|---|---|---|
| 16 | 11.56 | 0.15 | 0.072 |
| 128 | 27.7 | 0.31 | 0.6 |
| 256 | 30.96 | 13.8 | 0.92 |

In order to investigate the impact of the method of data filling in the shadow zone, we chose three different strategies; zero padding, linear interpolation in θ direction and using the Radon value with applying 3D DRT on the phantom. Due to the limited number points in the shadow zone, the method's impact is insignificant. In Table 6, CS phantom is considered with *nx=64* and *sx=32*.

Table 6
Comparing different methods in shadow zone filling

| Method of Filling Shadow Zone | PSNR |
|---|---|
| Zero Padding | 29.95 |
| Linear Interpolation | 30.96 |
| Radon Value Of Phantom | 30.99 |

To quantify the reconstruction image with FDK method, PSNR and SSIM of the whole slice and CNR of the ROI (yellow square in the Fig 22) of three phantoms were listed in Table 7. The valuation indexes of our proposed method were found to be close to FDK.

Variations in the $\mu_{max}$ and the MSE of the images result in a differences between PSNRs obtained from various phantoms.

Table 7
PSNR, SSIM and CNR of different phantoms by FDK and the proposed method

|  |  | Shepp-Logan | Head | CS |
|---|---|---|---|---|
| PSNR | FDK | 19.73 | 25.63 | 29.85 |
|  | The proposed method | 20.15 | 22.72 | 30.96 |
| SSIM | FDK | 0.49 | 0.90 | 0.93 |
|  | The proposed method | 0.46 | 0.62 | 0.92 |
| CNR | FDK | .03 | 0.12 | 25.5 |
|  | The proposed method | 0.44 | 0.41 | 13.8 |

### B. Visually Comparison

Compared to [4], Fig 20 (first row) shows that there is no artifact in the sinogram obtained from the proposed method.

While Lee et al [4] found three types of artifacts associated with the Grangeat's formula, which are respectively called *wrinkle*, *V-shaped* and *thorn* artifacts. In Fig 20, we compared the sinogram and reconstructed images of Lee's study with ours.

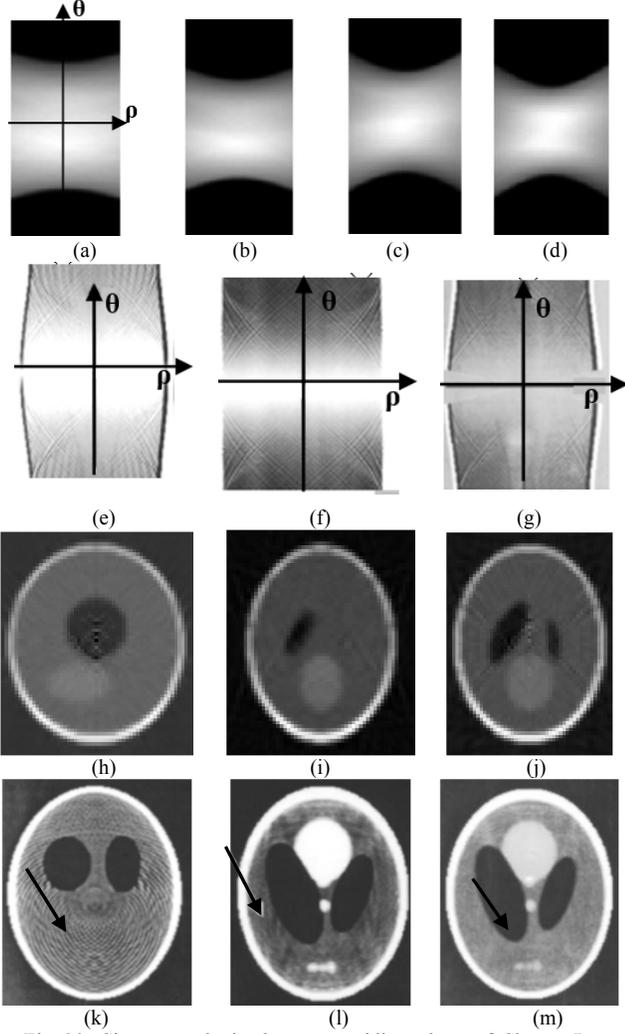

Fig 20. Sinogram obtained on a meridian plane of Shepp- Logan phantom of size 64×64×64 by our method. (a) $Radon_{[1]}(:,:,35)$, (b) $Radon_{[1]}(:,:,25)$, (c) $Radon_{[2]}(:,:,5)$, (d) $Radon_{[3]}(:,:,25)$, (e, f, g) Wrinkle, V-shaped and thorn artifacts in the second derivative Radon in [4] respectively, (h, i, j) Representative slices of the reconstructed image with the proposed method, (k, l, m) Reconstructed image with Wrinkle, V-shaped and thorn artifacts in [4] respectively.

In Fig 21, the values of the three randomized Radon diameters in the PPFT grid, related to the phantom No.1, Table 4, were shown. This figure is related to the output of block 9 in Table 3.

Two signals shown in Fig 21: $Radon_{[i]}(k,l,j)$, derived from applying 3D DRT on the original phantom (black line) and the Radon obtained from CBP by the proposed algorithm (red line), that $1 \leq i \leq 3$, $3 \times nx/2 \leq k \leq 3 \times nx/2$, $1 \leq l \leq nx+1$ and $1 \leq j \leq nx+1$.

In this numerical simulation, a represented slice of reconstructed objects introduced in section IV are shown in Fig 22. Comparing the original and reconstructed images makes clear that the proposed method effectively removed the three types of artifacts mentioned in [4].

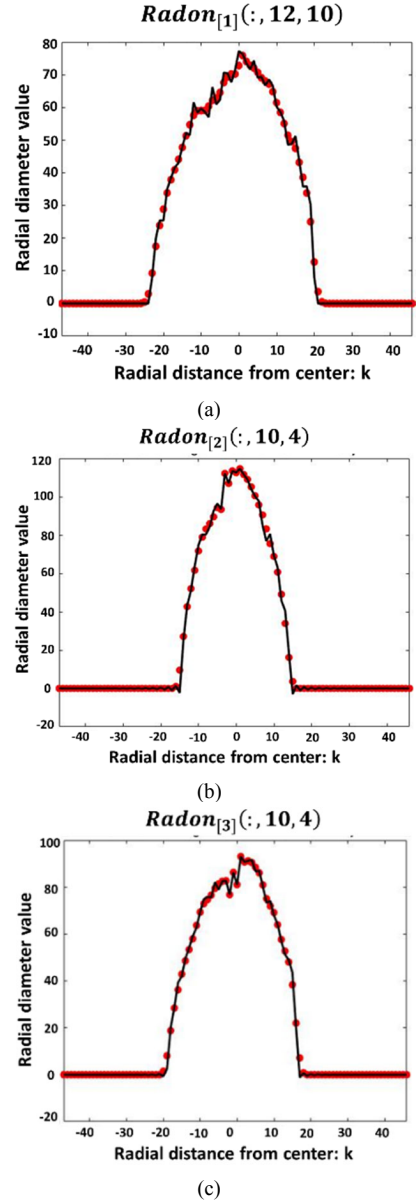

Fig 21. (a), (b) and (c) are three random Radon diameters correspond to $PP_1, PP_2$ and $PP_3$ respectively. The red signal is obtained by the proposed method, while the black one is obtained by applying 3D DRT to the phantom of No1, Table 4.

## C. Ability to Preserve Smooth Images and Edges.

Fig 23 demonstrates a representative slice of a noise-free image out of the CS phantom with size of $64 \times 64 \times 64$ voxels. The phantom (Fig 23) with a uniform background contains smoothly-changed intensity, such as octahedron in the upper-right corner and the sphere in the lower-left corner.

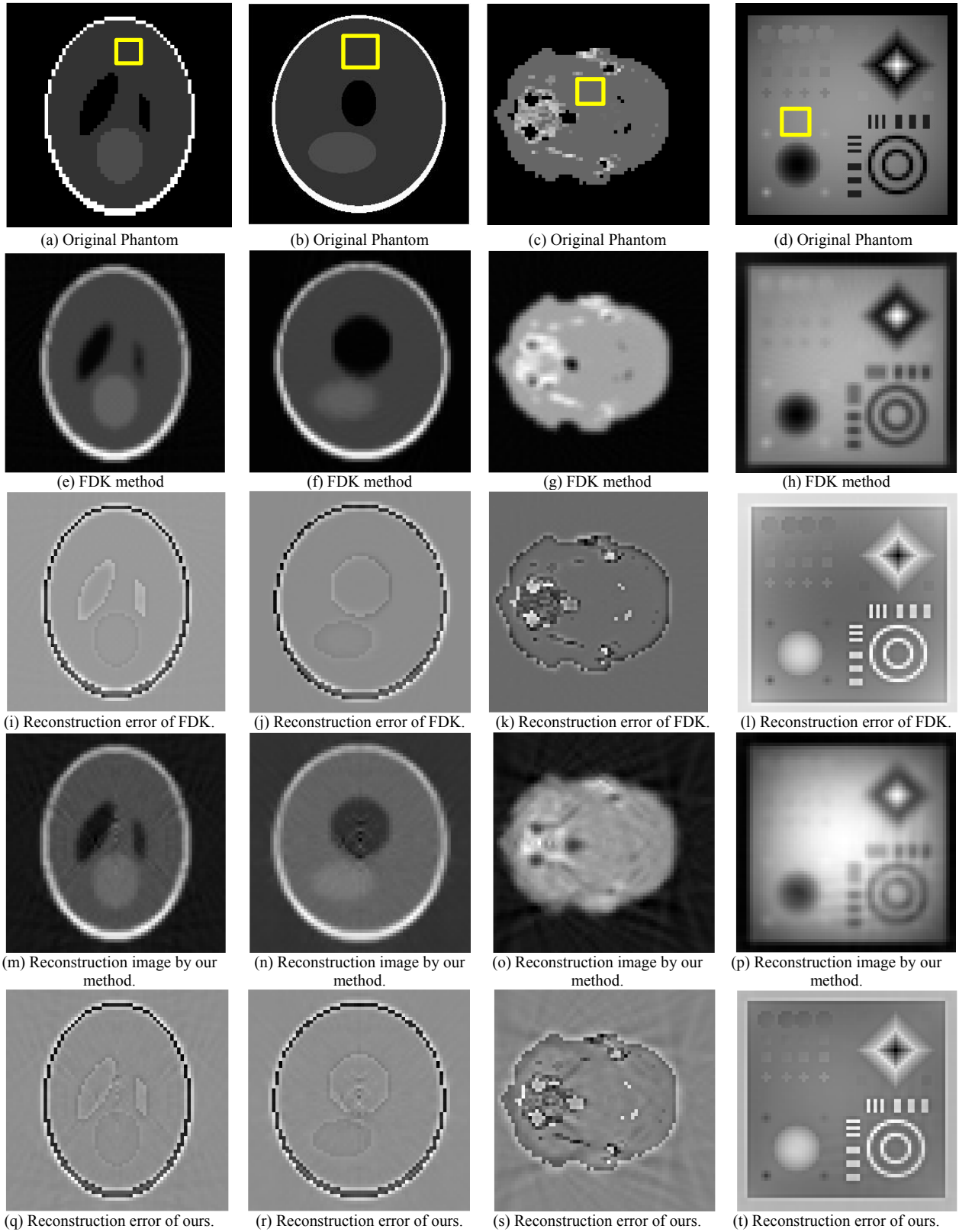

Fig 22. (a, c, d) Used three phantoms of size 64× 64 × 64. Yellow squares are selected as ROI to compare CNR. (a, b) 27[th] vertical slice, 28[th] transverse slice of Shepp-Logan [22], (c) 15[th] vertical slice of Head [24], (d) 32[nd] vertical slice of CS, (e, f, g, h) reconstructed image by FDK method, (i, j, k, l) Reconstruction errors by FDK, (m, n, o, p) reconstructed image by our method and (q, r, s, t) Reconstruction errors by ours.

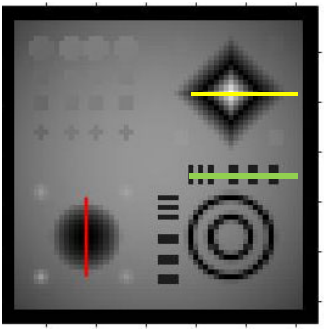

Fig 23. A representative slice of the original CS phantom 64×64×64.

Fig 24.a and b show the comparative profiles of the phantom along the center of sphere and octahedron includes the red and yellow lines in Fig 23 respectively. The intensity curves along these lines exhibited numerous small and unnatural constant intensity artifacts in the reconstructed image as well as FDK indicating its capacity to preserve smooth images.

Furthermore, the phantom contains a set of line objects to measure the resolution.

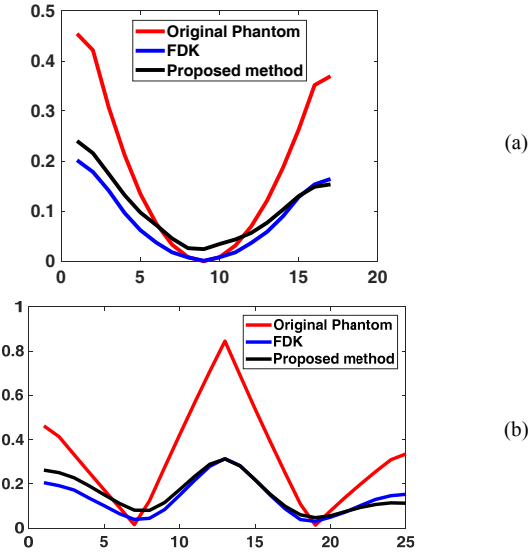

Fig 24. Profiles through the red and yellow line in Fig 23; (a) Comparative plot of FDK and our approach along red line; (b) Comparative plot of FDK and our approach along yellow line

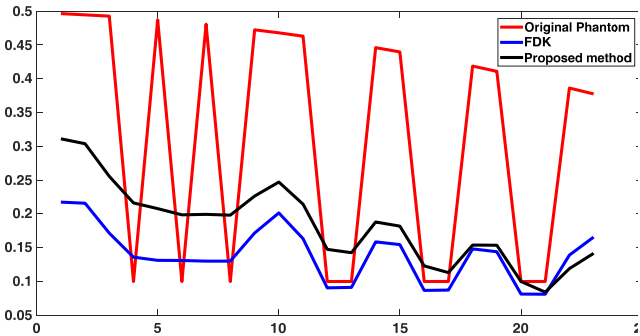

Fig 25. Profile through the barcode (green line in Fig 23) related to reconstructed images by FDK and our method

To evaluate our method in the edge area, the strip area (Barcode) in Fig 23 was selected. Fig 25 shows that the proposed method preserved the edges resembling FDK.

*D. Ability to Preserve Local Structure*

We used SSIM in our study to measure the degree of similarity in local structures between reconstructed images and the original image.

Fig 26 shows that the SSIM map of the Shepp-Logan phantom in our method was very close to FDK, indicating a good ability in preserving structures.

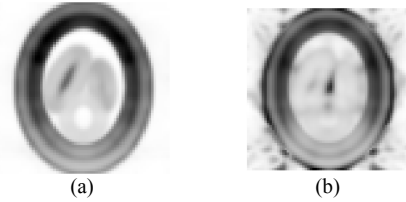

(a)    (b)

Fig 26. SSIM map of the head phantom using different methods. (a) FDK (SSIM=0.85), (b) Ours (SSIM=0.82)

*E. Reconstructed Images with Noisy Projections.*

In the case of normal clinical exposures, assuming monochromatic source, the X-ray CT measurements are often modeled as the sum of a Poisson distribution representing photon counting statistics and an independent Gaussian distribution representing additive electronic noise [26], i.e.,

$$z_i = Poisson(I_i exp(-\phi f_i)) + Gaussian(m_i, \sigma_i) \quad (56)$$

Where $f_i$ is the attenuation coefficient map, $z_i$ denotes the number of X-ray photons incident on detector along the $i^{th}$ X-ray path, and $I_i$ is the blank scan factor, $m_i$ and $\sigma_i$ indicate the mean and standard deviation of electronic noise that has been converted to photon units. The offset mean *m* of background signals such as dark current can be estimated using blank measurements prior to each scan and subtracted from the measured intensity, so we assume $m_i=0$ hereafter. In this section we assume the standard deviation is 20.

We generally do not expect analytic methods to eliminate noises. The analytical methods are somehow just a mapping of the detectors values to the desired image and generally we do not expect analytic methods to eliminate noise. However, such methods are a means of producing an initial exact image to initiate iterative reconstruction (IR) methods, then IR methods will remove the noise and artifacts.

To evaluate the behavior of our method in noise removal, we have added the Gaussian noise to the projections. After noise addition (Fig 27.c), PSNR of reconstructed image decreased from 20.15 dB to 19.98 dB. Also the CNR of the ROI (Fig 22. a) reduced from 0.44 to 0.24. So, the results show that our proposed method have no ability to remove the projection noises.

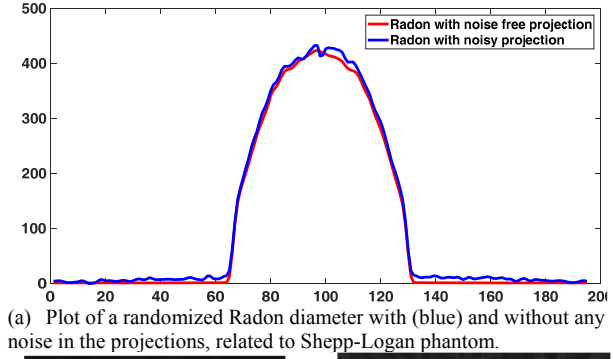
(a) Plot of a randomized Radon diameter with (blue) and without any noise in the projections, related to Shepp-Logan phantom.

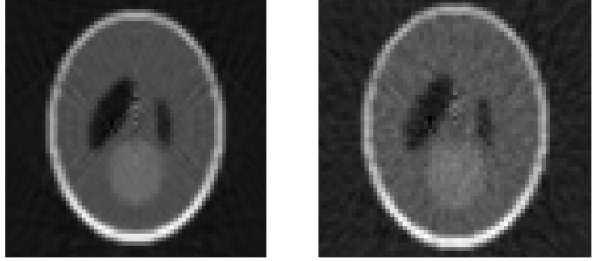

(b) Reconstructed images from noise-free projections.

(c) Reconstructed images from noisy projections.

Fig 27. Comparative results of noise-free and noisy projections

In the next section, we explore the ability of the proposed method in generating an image to be useful for the initialization of the iterative algorithms in low-dose CBCT reconstruction.

*F. Reconstruction of High-quality Images of Sparse-View Angle CBCT Using the Image Reconstructed by the Proposed Method as an Initial Image of the Iterative Algorithm*

We have used in all sections of this paper 360 projections, which was good enough for the reconstruction of a high-quality image. Now, we want to reconstruct high-quality images of sparse-view angle CBCT using 120 projections around 360°. One of the most successful methods of reconstructing a low-dose CBCT with a limited- number of projections is Compression Sensing (CS). However, the first step in achieving a good image is to produce an accurate Radon space. The Radon space provides the basis for the use of Fourier-based iterative methods.

The tomography operator $\phi$ in $y = \phi f + w$, is ill-posed and therefore cannot be inverted. We apply variational reconstruction methods wherein a solution is found through the convex optimization problem

$$f^* = \arg\min_f \frac{1}{2}\|\phi f - y\|^2 + \lambda J(f) \tag{57}$$

Where y is the measurement and $J(f)$ is a prior energy, whose choice will be explained in the following sections.

According to FST, we explore a FFT version of $\phi$ as it corresponds to the sampled FFT of the image along the discretized rays.

In this study, due to the limited number of projections, there are only a few non-zero diameters in the 3D PPFT Radon space. In other words, based on FST, and due to the limited view angle conditions in CBCT, the generated 3D Radon space includes a partial Radon transform or integral of 3D object along a limited number of equispaced rays at spherical orientations corresponding to the lines of the 3D-PP grid.

Recovering from tomography by $y = \phi f$ is equivalent to inpainting the missing Fourier frequencies. We assume the partial noisy Fourier measures as

$$\forall \omega \in \Omega, \hat{y}[\omega] = \hat{f}[\omega] + \hat{w}[\omega] \tag{58}$$

Where $\hat{y}$ is 1D radial FFT of the 3D Radon space, $\hat{f}$ denotes 3D PPFT of 3D unknown image and $\hat{w}[\omega]$ is the measured noise on projections in the frequency domain.

Iterative reconstruction algorithms have demonstrated an excellent performance in improving the quality of the image. We therefore use an optimization problem to increase the quality of reconstructed images. The total variation regularization is as follows:

$$f^* = \arg\min_f \frac{1}{2}\sum_{\omega \in \Omega}\left|\hat{f}[\omega] - \hat{y}[\omega]\right|^2 + \lambda\|f\|_{TV} \tag{59}$$

The reconstruction primary results indicate the FDK algorithm cannot reduce the noise and therefore significant artifacts are created. The reconstructed images with TV look more natural and smoother, emphasizing the ability of TV in preserving smooth regions as well as suppressing the noise

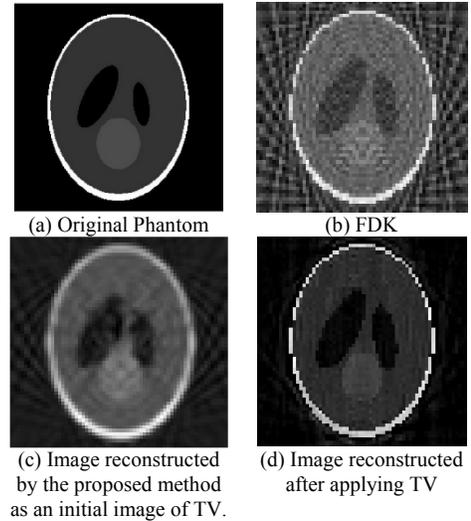

(a) Original Phantom    (b) FDK

(c) Image reconstructed by the proposed method as an initial image of TV.

(d) Image reconstructed after applying TV

Fig 28. (a) A representative slice of the original Shepp- Logan phantom; and reconstructed images by different reconstruction algorithms with 120 projections (sparse view- angle).

## VI. DISCUSSION

Although the analytical formulations of exact reconstruction methods are theoretically error free, there are some limitations that make the use of approximate methods more welcome [27]. The first category pertains to the noise that appears in reconstructed images. Owing to the method presented in this article, this type of noises reported in previous papers has disappeared. Three types of artifacts associated with the Grangeat's formula were reported, known as thorn, wrinkle, and V-shaped artifacts [4].

The second limitation of these algorithms is consistency condition. A set of conditions is full if the conditions are both necessary and sufficient. Accordingly once the objects are exactly rebuilt, there is a perfect condition. As an example, one of these conditions states that each plane of the intersected object should have at least a point on the source path [2]. If so, it would be possible to achieve an accurately reconstructed object, while there is no such status in the conventional CT with a circular source path. Furthermore, considering the risk of cancer due to x-ray exposes, researchers have moved on to propose ways to produce a precise, high-quality image while reducing radiation with a limited number of projection data. Obviously, by reducing the number of projections due to lower risk of x-ray, there is no full data requirement for an exact reconstruction.

The third set of problems is related to low efficiency on the removal of noises from the projections. These methods treat noise-related problems as a supplement because they ignore noise measurement in the problem formulation [27] and reconstructed objects are just a weighted mapping of projections value.

In order to remove the two last classes of artifacts, it must be mentioned that the combination of exact reconstruction and iterative algorithms can result in the production of optimal image quality in sparse view angles and noise treatment conditions. One of the successful methods with these features is FIRM which resembles Equally Sloped Tomography (EST). So far, different articles have been presented in this area for Fan Beam CTs, all of which having generated 2D Radon space from fan beam projections, applied Fourier transforms and expressed a regularization function [28]–[31]. In this paper, we have sought to provide an accurate 3D Radon space with 3D PPFT features to facilitate the use of FIRM in Cone Beam CTs. The proposed method is also useful for developing intuition, and also for initializing iterative algorithms associated with statistical reconstruction methods.

One of the other aims is to reduce the computational complexity, i.e. to produce a faster method. The most taxing procedure during the first phase is computation of line-integrals in the detector plane. By applying the direct Fourier method in reverse for this computation, we reduce the complexity of phase 1 from $O(N^4)$ to $O(N \log N^3)$. The second Phase is performed via direct Fourier methods which reduces the complexity of $O(N^4)$ to $O(N \log N^3)$ as well.

Our code, Cone2Radon toolbox, is available open source. After the acceptance of the paper, we will post it on https://misp.mui.ac.ir/en.

## VII. CONCLUSION AND FUTURE WORK

In this paper, we have shown 2D and 3D DRT have been used for an accurate modeling of continuum in parallel with the increasing number of samples. Meantime, we demonstrated that inverse DRT could be used in reconstruction from CBP using equispaced detectors.

Furthermore, we provide 3D Radon space in linogram fashion to allow the use of Fourier based methods with CBP for the reconstruction of 3D images in a sparse view angle CBCT with no error of interpolation.

The recent use of FIRM has rendered possible achieving high-quality two-dimensional (2D) images from a fan beam CT with a limited number of projections. This technology reduces the dosage of radiation to which a patient is exposed, thereby reducing the risk of cancer. In the future, our studies will focus on the use of Fourier based Compressive Sensing (CS) algorithm for CBCT image reconstruction in sparse view angles conditions.

Analytical reconstruction in CT is very well-suited for parallel implementations on GPUs. At the interpolation stage, the code for the proposed algorithm is very time consuming. For this reason, we will intend to parallel implementation of this approach on GPU in the future.

ACKNOWLEDGMENT

We would like to thank Dr Sayyed Masoud Hashemi for his valuable guidance. This work is sponsored in part by Behyaar Sanat Sepahan Co.